\documentclass[%
reprint,
amsmath,amssymb,
aps,
pra,
]{revtex4-2}
\usepackage{graphicx}
\usepackage{dcolumn}
\usepackage{bm}
\usepackage[colorlinks=true,linkcolor=blue,citecolor=blue,urlcolor=blue]{hyperref}
\usepackage{xcolor}
\usepackage{physics}
\usepackage{multirow}
\usepackage{mathdots}
\begin{document}
	\title{Chiral Quantum Transport with Perfect Circulation: From Floquet Engineering to Anyonic Dynamics}
	
	\author{Chaorong Guo}
	\affiliation{Zhejiang Key Laboratory of Quantum State Control and Optical Field Manipulation, Department of Physics, Zhejiang Sci-Tech University, Hangzhou 310018, China}
	
	\author{Hongzheng Wu}
	\affiliation{Zhejiang Key Laboratory of Quantum State Control and Optical Field Manipulation, Department of Physics, Zhejiang Sci-Tech University, Hangzhou 310018, China}
	
	\author{Zenong Zhou}
	\affiliation{Zhejiang Key Laboratory of Quantum State Control and Optical Field Manipulation, Department of Physics, Zhejiang Sci-Tech University, Hangzhou 310018, China}
	
	\author{Ai-Xi Chen}
	\affiliation{Zhejiang Key Laboratory of Quantum State Control and Optical Field Manipulation, Department of Physics, Zhejiang Sci-Tech University, Hangzhou 310018, China}
	
	\author{Xiaobing Luo}
	\email{xiaobingluo2013@aliyun.com}
	\affiliation{Zhejiang Key Laboratory of Quantum State Control and Optical Field Manipulation, Department of Physics, Zhejiang Sci-Tech University, Hangzhou 310018, China}
	\date{\today}
	
	\begin{abstract}
		Perfect chiral circulation---the sequential transfer of a quantum state around a closed loop with unit fidelity---has been achieved in specific few-site systems, yet the universal physical conditions underlying this phenomenon remain unclear. We prove that discrete translational invariance and an equidistant energy spectrum together constitute the necessary and sufficient conditions for perfect chiral circulation. With this criterion established, an exact closed-form Hamiltonian valid for arbitrary $N$-site rings naturally follows. In the minimal three-site ring, we demonstrate two physically distinct realizations: Floquet engineering of a driven open chain that restores translational invariance by equalizing the couplings, and correlated doublon dynamics in an anyon-Hubbard model where fractional statistics intrinsically provide the chiral flux that renders the spectrum equidistant. Our results establish unified physical criteria for perfect chiral circulation and demonstrate their applicability across diverse platforms such as superconducting circuits, cold atoms, classical electrical circuits, and photonic synthetic dimensions.
	\end{abstract}
	\maketitle
	
	\section{Introduction}
	Directed quantum transport is a central theme in modern quantum physics, pursued across a variety of platforms~\cite{cooper2019topological, goldman2016topological, Ozawa2019, Blais2021}. The quantum ratchet~\cite{PhysRevLett.84.2358,Denisov2002,RevModPhys.81.387,reimann2002brownian} achieves rectified motion by breaking relevant space-time symmetries, enabling directional transport of particles or energy~\cite{PhysRevLett.131.133401}. Topological pumping~\cite{PhysRevB.27.6083,Citro2023} provides another route to robust quantized directed transport, where the transport distance per cycle is determined by the bulk topological invariant (the Chern number)~\cite{TKNN1982}. Since matter-wave solitons can achieve high-fidelity quantum state transfer during propagation, the quantum ratchet effect and Thouless topological pumping of solitons have also attracted widespread attention~\cite{FuYe2022PRL,HuXX2024}.
	
	On ring geometries, synthetic gauge fields break time-reversal symmetry, giving rise to unidirectional chiral transport. Most notably, restricted to the minimal three-site geometry, perfect chiral circulation of single-particle excitations enables high-fidelity state transfer and has been demonstrated in a variety of systems~\cite{Roushan2017,Koch2010,Wang2019,Downing2020,Herrmann2022,PhysRevLett.126.123603,Qi2022}. Providing an important initial step toward scalable chiral networks, a recent study~\cite{PhysRevApplied.23.054080} extended perfect chiral transport to $N$-node ring networks. However, the scheme necessitates auxiliary nodes, and the fundamental physics enabling this phenomenon for arbitrary rings has yet to be fully elucidated. Rather than relying on structural additions, a more fundamental route lies in controlling the spectral properties of the system. As established in linear chains, engineering the energy spectrum is the key to unlocking perfect end-to-end state transfer~\cite{bose2003quantum, christandl2004perfect, Kay2010}. This shift in perspective leads to a fundamental question: Can we establish a universal criterion, rooted in spectral structure, to realize perfect chiral excitation circulation on a ring?
	
	In this paper, we develop an analytical framework for perfect chiral circulation of single excitations in $N$-site rings of arbitrary size, following the sequence $|1\rangle \to |2\rangle \to \cdots \to |N\rangle \to |1\rangle$. We identify discrete translational invariance and an equidistant energy spectrum as the two universal, necessary and sufficient criteria for this phenomenon. Uniquely determined by these criteria, the exact closed-form Hamiltonian—featuring long-range hopping and chiral phases—naturally follows for any $N$. To explicitly validate the universality of these criteria, we examine the minimal three-site ring as a testbed, presenting two physically distinct realizations: Floquet engineering in an open-boundary chain, and correlated doublon hopping in an anyon-Hubbard model. Despite their profound physical differences, both schemes yield the same target Hamiltonian, underscoring that our two criteria dictate perfect chiral circulation independently of any specific platform.

	\begin{figure}[tb]
		\centering
		\includegraphics[width=\linewidth]{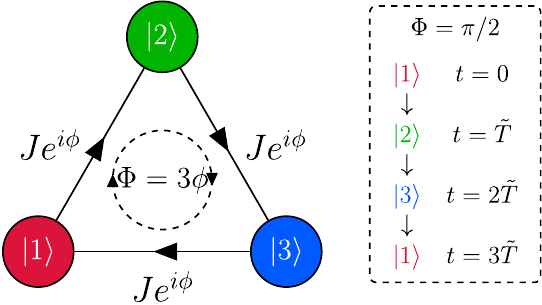}
		\caption{\label{fig:schematic} 
			Minimal chiral ring ($N=3$). (Left) Triangular ring model with uniform hopping amplitude $J$ and hopping phase $\phi$ per bond, yielding a total synthetic flux of $\Phi = 3\phi$. (Right) The chiral excitation circulation is indicated by text and arrows as $|1\rangle \to |2\rangle \to |3\rangle \to |1\rangle$ at $\Phi = \pi/2$.
		}
	\end{figure}
	
	\section{UNIVERSAL CRITERIA AND TARGET HAMILTONIAN}
	We consider a particle on an $N$-site ring with site states $|n\rangle$ ($n = 1, 2, \dots, N$), governed by a translationally invariant Hamiltonian $\hat{H}$. Perfect chiral transport requires that after a single-step evolution time $\tilde{T}$, the propagator $\hat{U}(\tilde{T}) = e^{-i\hat{H}\tilde{T}}$ acts as a cyclic shift: $\hat{U}(\tilde{T})|n\rangle = |n+1\rangle$ for all $n$ ($|N+1\rangle \equiv |1\rangle$), up to irrelevant phases (see Appendix~\ref{sec:S1}). The matrix elements $U_{mn} = \langle m|\hat{U}(\tilde{T})|n\rangle = \delta_{m,n+1}$ form a cyclic permutation matrix. Diagonalizing $\hat{U}(\tilde{T})$ yields eigenvalues equally spaced on the unit circle, which \emph{requires} the energy spectrum to form an arithmetic progression (see Appendix~\ref{sec:S1}):
	\begin{equation}
		E_k = \frac{2\pi k}{N\tilde{T}}, \quad 
		k = 0,1,\dots,N-1.
		\label{eq:spectrum}
	\end{equation}
	Translational invariance guarantees that the 
	eigenstates are the discrete Bloch modes 
	\begin{equation}
		|k\rangle = \frac{1}{\sqrt{N}} \sum_{n=1}^{N} 
		e^{i\frac{2\pi kn}{N}} |n\rangle.
		\label{eq:bloch}
	\end{equation}
	Acting with $\hat{U}(\tilde{T}) = e^{-i\hat{H}\tilde{T}}$ on a site state and substituting $E_k \tilde{T} = 2\pi k/N$, one obtains
	\begin{equation}
		\begin{split}
			\hat{U}(\tilde{T})|n\rangle 
			&= \frac{1}{\sqrt{N}} \sum_{k=0}^{N-1} 
			e^{-i\frac{2\pi kn}{N}}\, e^{-iE_k \tilde{T}} |k\rangle \\
			&= \frac{1}{\sqrt{N}} \sum_{k=0}^{N-1} 
			e^{-i\frac{2\pi k(n+1)}{N}} |k\rangle 
			= |n\!+\!1\rangle,
		\end{split}
		\label{eq:U_circulation}
	\end{equation}
	confirming that translational invariance combined with an equidistant spectrum is sufficient for perfect chiral circulation (see Appendix~\ref{sec:S2}). Under these sufficient conditions, the Hamiltonian matrix elements in the site basis naturally follow as 
	\begin{equation}
		\begin{split}
			\hat{H}_{mn} &= \sum_{k=0}^{N-1} E_k 
			\langle m|k\rangle \langle k|n\rangle \\
			&= \frac{2\pi}{N\tilde{T}} \cdot 
			\frac{1}{e^{i\frac{2\pi(m-n)}{N}} - 1}, 
			\quad (m \neq n),
		\end{split}
		\label{eq:Hmn}
	\end{equation}
	where the diagonal elements are a uniform constant 
	absorbed into the energy zero point. The Hamiltonian can be written as (see Appendix~\ref{sec:S2})
	\begin{equation}
		\hat{H} = -\sum_{m > n} \left( |J_{mn}|\,
		e^{i \Phi_{mn}}\, \hat{a}^\dagger_m \hat{a}_n
		+ \text{H.c.} \right),
		\label{eq:H_ideal}
	\end{equation}
	with
	\begin{subequations}
		\begin{align}
			\frac{|J_{mn}|}{J} &=
			\frac{\left|\sin\left(
				\frac{\pi}{N}\right)\right|}
			{\left|\sin\left(\frac{\pi}{N}(m-n)
				\right)\right|},
			\label{eq:amplitude} \\
			\Phi_{mn} &= -\frac{\pi}{N}(m-n)
			+ \frac{\pi}{2},
			\quad (m > n), \label{eq:phase}
		\end{align}
	\end{subequations}
	where $J \equiv |J_{\text{NN}}|$ is the
	nearest-neighbor hopping amplitude,
	$\hat{a}^\dagger_m$ ($\hat{a}_n$) creates
	(annihilates) a particle at site $m$ ($n$), and
	site indices are periodic with $|N+1\rangle \equiv |1\rangle$. The chiral phases $\Phi_{mn}$ break time-reversal
	symmetry. The hopping phases are gauge-dependent,
	but the total magnetic flux enclosed by any closed
	loop is gauge-invariant and uniquely determines
	the dynamics. In particular, the 
	distance-dependent part of 
	Eq.~\eqref{eq:phase} can be removed by a gauge 
	transformation 
	(see Appendix~\ref{sec:S2}),
	so that every pairwise coupling phase reduces to 
	$\Phi_{mn} = \pi/2$.
	
	The Hamiltonian governing the backward circulation $|n\rangle \to |n\!-\!1\rangle$ is precisely the complex conjugate of the forward one (see Appendix~\ref{sec:S3}); reversing all phases therefore reverses the circulation direction.
	
\begin{figure}[tb]
		\centering
		\includegraphics[width=\linewidth]{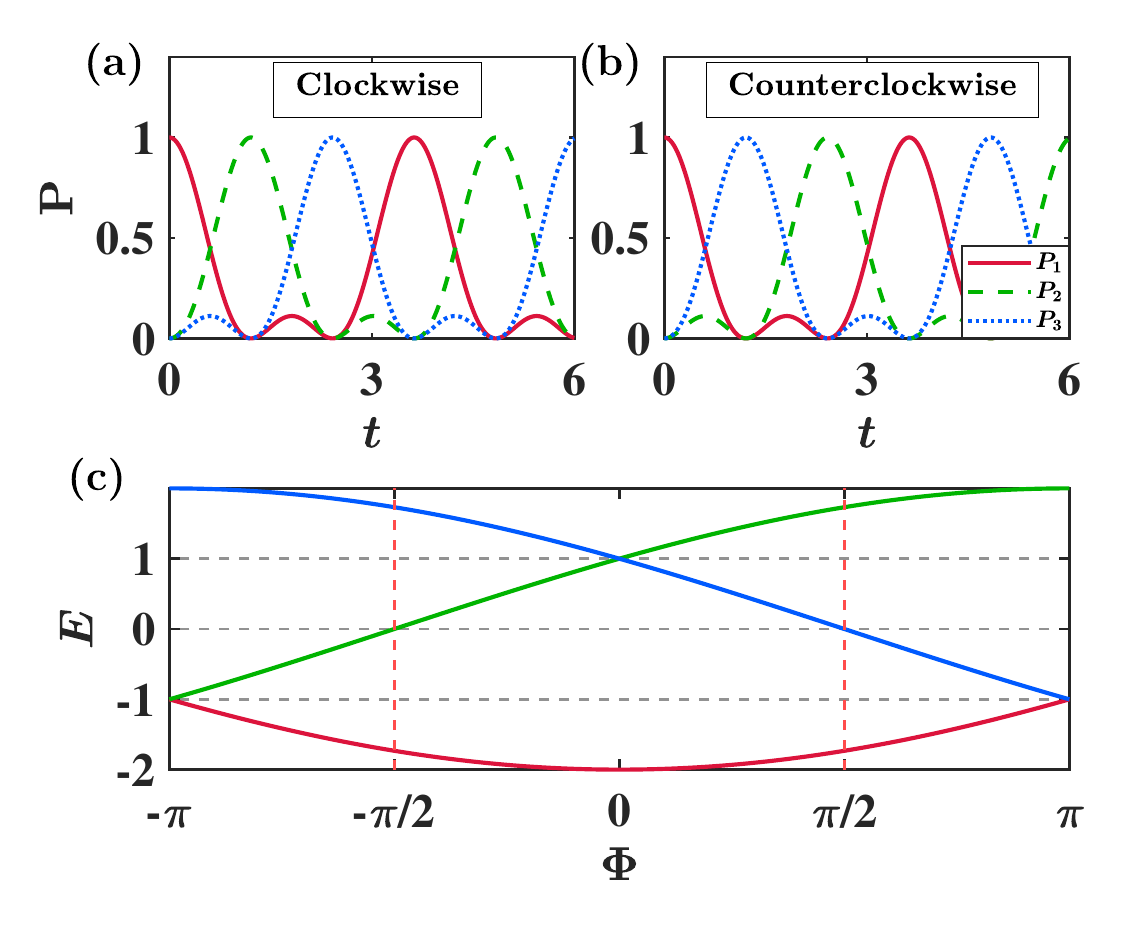}
		\caption{\label{fig:N3_dynamics} 
			Dynamics and spectrum for $N=3$.  
			(a) Clockwise circulation 
			($\Phi = \pi/2$). Site populations 
			$P_n(t) = |\langle n|\psi(t)\rangle|^2$ for 
			$n=1$ (red), $2$ (green), and $3$ (blue).
			(b) Counterclockwise circulation 
			($\Phi = -\pi/2$).
			(c) Eigenenergy spectrum versus 
			flux $\Phi$. Dashed lines mark 
			$\Phi = \pm\pi/2$, where the level spacing 
			becomes uniform. The initial state is $|1\rangle$, and time is in units of $1/J$.
		}
	\end{figure}

	For the minimal case $N=3$ (Fig.~\ref{fig:schematic}), Eqs.~\eqref{eq:amplitude} and~\eqref{eq:phase} reduce to a uniform hopping amplitude $J$ and a hopping phase $\phi = \pi/6$ per bond, yielding a total flux $\Phi = 3\phi = \pi/2$. Figure~\ref{fig:N3_dynamics}(c) confirms that this flux renders the spectrum equidistant, and the resulting dynamics exhibit perfect clockwise circulation $|1\rangle \to |2\rangle \to |3\rangle \to |1\rangle$ [Fig.~\ref{fig:N3_dynamics}(a)]. Applying the opposite flux $\Phi = -\pi/2$ reverses the circulation direction while leaving the spectrum unchanged [Figs.~\ref{fig:N3_dynamics}(b) and (c)].
	
	\begin{figure}[tb]
		\centering
		\includegraphics[width=\linewidth]{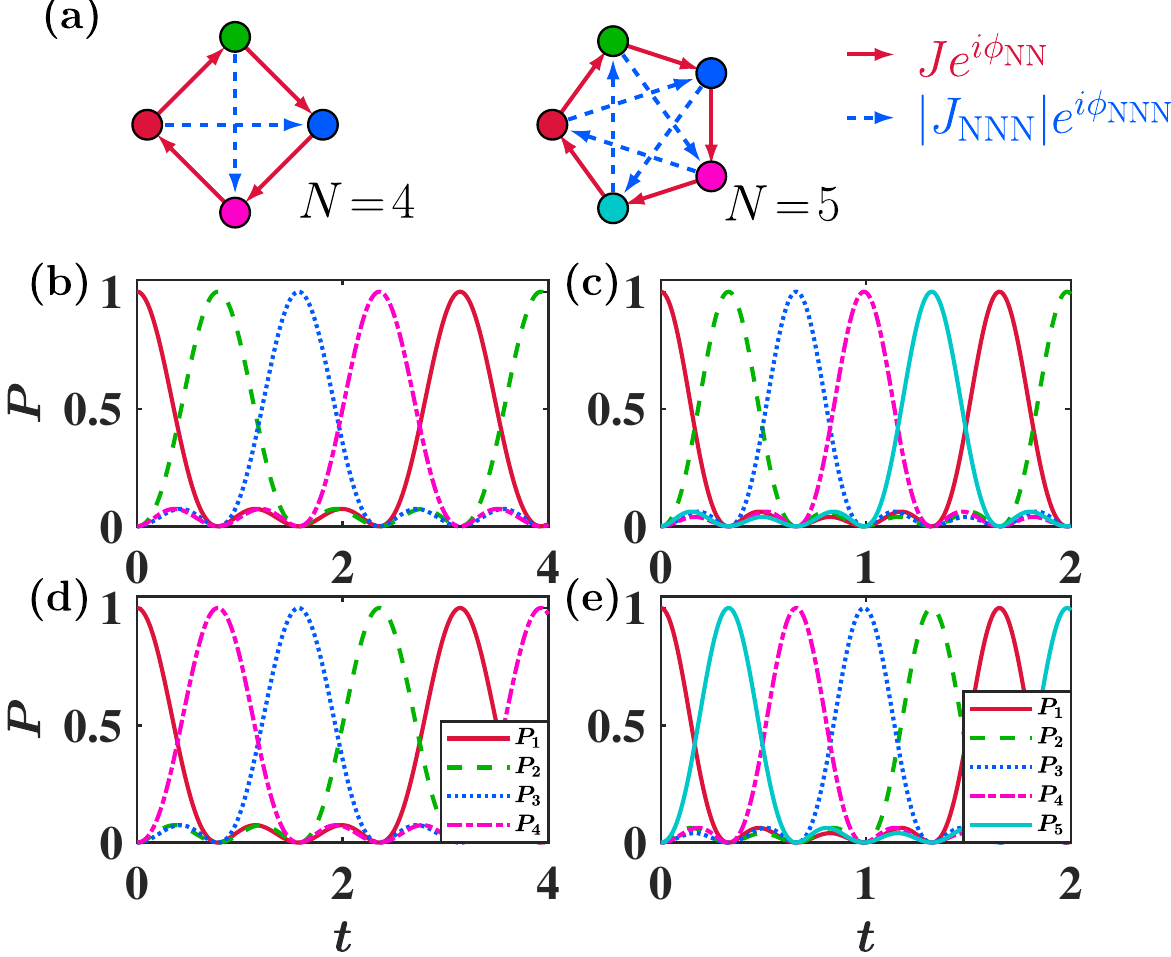}
		\caption{\label{fig:scalable} 
			Scalable chiral dynamics ($N=4, 5$). (a) Connectivity schematics for rings with $N=4$ and $N=5$. Red solid arrows: NN hopping; blue dashed arrows: NNN hopping. For $N=4$: $J/J_{\text{NNN}} = \sqrt{2}$, $\Phi_{\text{NN}} = \pi/4$, $\Phi_{\text{NNN}} = 0$. For $N=5$: $J/J_{\text{NNN}} = (1+\sqrt{5})/2$, $\Phi_{\text{NN}} = 3\pi/10$, $\Phi_{\text{NNN}} = \pi/10$. (b, c) Clockwise circulation for $N=4$ and $N=5$. (d, e) Counterclockwise circulation for $N=4$ and $N=5$. Hopping amplitudes and phases follow Eqs.~\eqref{eq:amplitude} and~\eqref{eq:phase}. The initial state is $|1\rangle$, and time is in units of $1/J$.
		}
	\end{figure}

	\begin{figure}[tb]
		\centering
		\includegraphics[width=\linewidth]{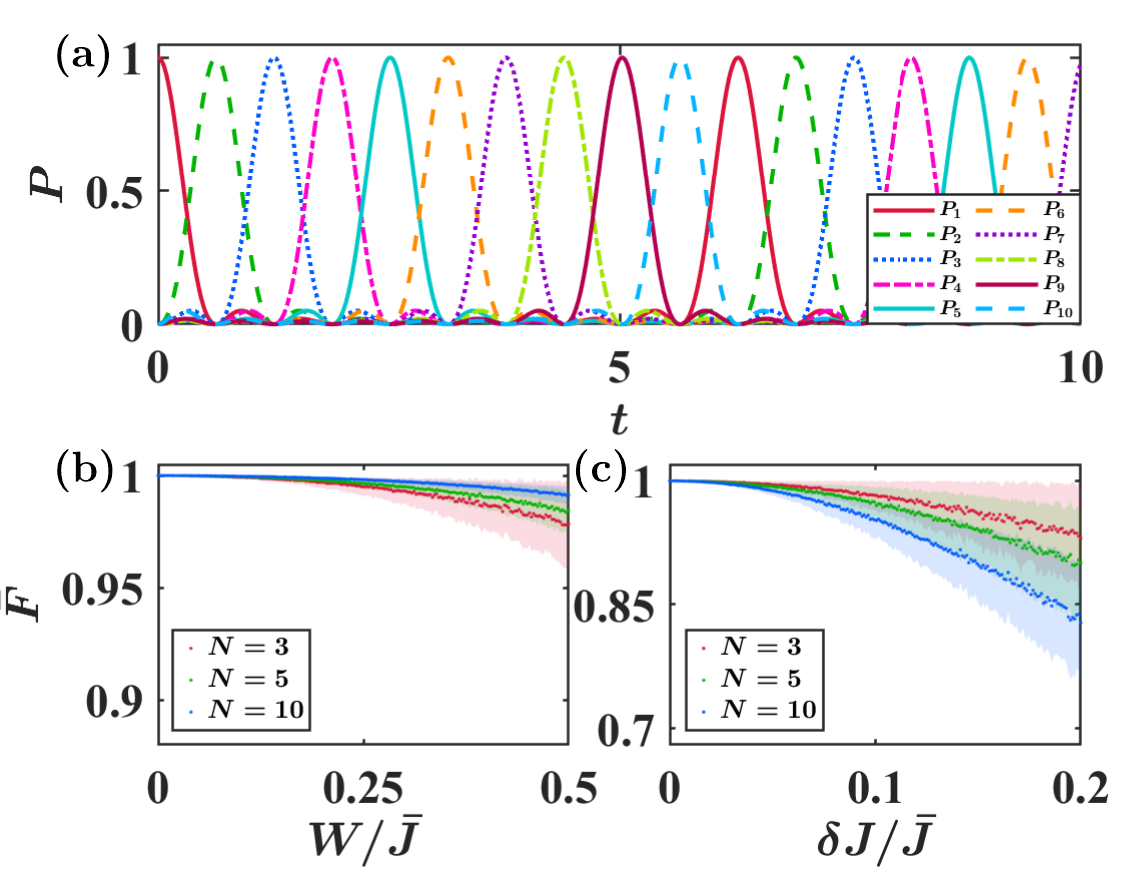}
		\caption{\label{fig:robust}
			Scalability and robustness.
			(a) Chiral circulation for 
			$N=10$. The Hamiltonian includes all-to-all 
			couplings given by 
			Eqs.~\eqref{eq:amplitude} 
			and~\eqref{eq:phase}.
			(b) Average transfer fidelity $\bar{F}$ versus on-site disorder $W/\bar{J}$ for $N=3$, $5$, $10$. 
			(c) $\bar{F}$ versus hopping disorder $\delta J/\bar{J}$.
			Here, $W$ and $\delta J$ denote the maximum on-site and hopping disorder strengths. For each site and bond, the disorder is randomly and uniformly sampled from $[-W, W]$ and $[-\delta J, \delta J]$, respectively, and scaled by the mean hopping amplitude $\bar{J}$. 
			Results are averaged over 300 disorder realizations; shaded regions show the sample-to-sample spread.
			The initial state is $|1\rangle$, and time is in units of $1/J$.
		}
	\end{figure}
	
	\section{Scalability and robustness}
	For $N > 3$, the all-to-all Hamiltonian of Eq.~\eqref{eq:H_ideal} extends beyond nearest-neighbor couplings, as illustrated in Fig.~\ref{fig:scalable} for $N=4$ and $N=5$. Specifically, Eqs.~\eqref{eq:amplitude} and~\eqref{eq:phase} give rise to a network of nearest-neighbor (NN) and next-nearest-neighbor (NNN) hopping [Fig.~\ref{fig:scalable}(a)], with $J/J_{\text{NNN}} = \sqrt{2}$ for $N=4$ and $(1+\sqrt{5})/2$ (the golden ratio) for $N=5$ (see Appendix~\ref{sec:S4}). The resulting dynamics confirm unit-fidelity chiral circulation in both directions for both system sizes [clockwise in Figs.~\ref{fig:scalable}(b) and (c) for $N=4$ and $5$, respectively; counterclockwise in Figs.~\ref{fig:scalable}(d) and (e)].
	
	Consistent with the analytical construction, for $N=10$ [Fig.~\ref{fig:robust}(a)], the site populations transfer sequentially through all ten sites with unit fidelity. To probe robustness, we introduce random on-site energies $\epsilon_n \in [-W, W]$ and hopping fluctuations $\delta J_{mn} \in [-\delta J, \delta J]$ [Figs.~\ref{fig:robust}(b) and (c)], keeping the hopping phases fixed. We quantify this by the average transfer fidelity
	\begin{equation}
		\bar{F} = \frac{1}{N}\sum_{n=1}^{N} 
		\left|\langle n\!+\!1|\hat{U}(n\tilde{T})
		|1\rangle\right|^2,
		\label{eq:fidelity}
	\end{equation}
	i.e., the mean probability of reaching the target site $|n\!+\!1\rangle$ after $n$ steps.
	With disorder strengths normalized by the mean hopping amplitude $\bar{J} = \frac{2}{N(N-1)}\sum_{m>n}|J_{mn}|$, $\bar{F}$ averaged over 300 disorder realizations remains above $0.95$ for on-site disorder $W/\bar{J} \lesssim 0.5$, a tolerance that improves for larger rings. For hopping disorder $\delta J/\bar{J} \lesssim 0.1$, $\bar{F} > 0.9$. However, larger rings suffer from increased sensitivity due to error accumulation across the $N$ sequential steps.
	
	\begin{figure}[tb]
		\centering
		\includegraphics[width=\linewidth]{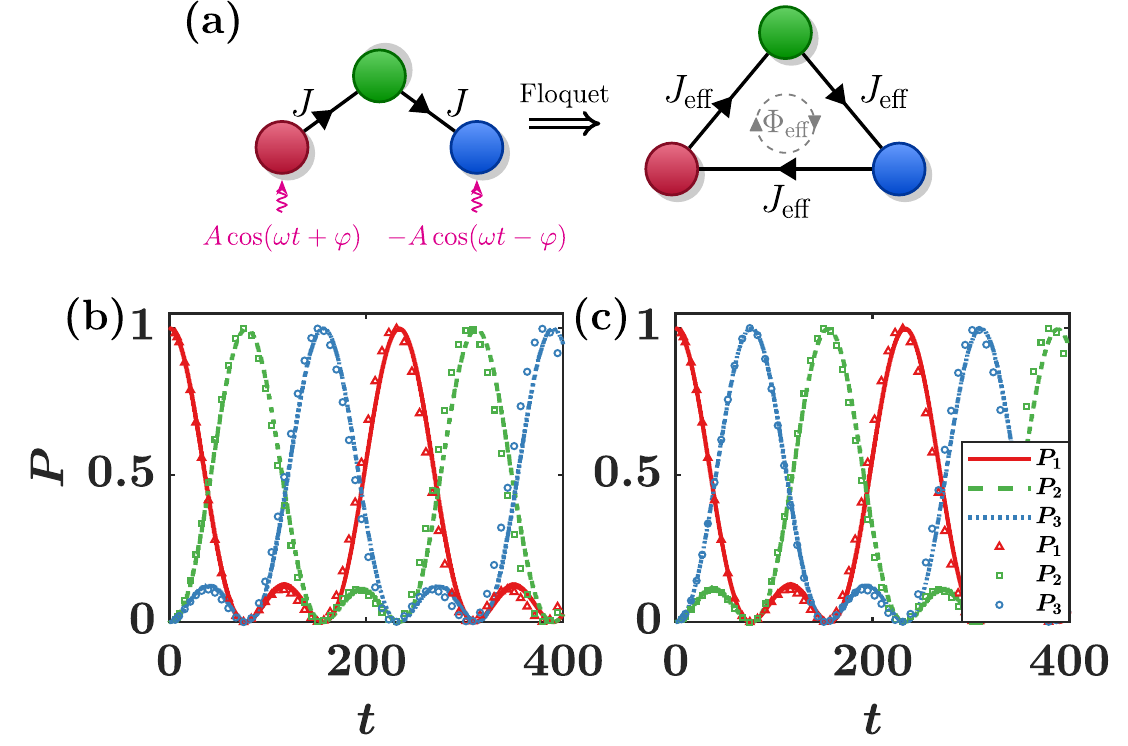}
		\caption{\label{fig:floquet}   
			(a) Floquet synthesis of chiral transport. 
			Asymmetric phase-shifted periodic driving 
			converts an open three-site chain (left) into 
			an effective triangular model with uniform 
			coupling $J_{\text{eff}}$ and chiral flux 
			$\Phi_{\text{eff}}$ (right).
			(b) Clockwise circulation 
			($\varphi = \pi/3$). Solid curves show the 
			exact numerical results from the driven 
			Hamiltonian~\eqref{eq:H_driven}; markers show 
			the analytical results from 
			$\hat{H}_{\text{eff}}$ in 
			Eq.~\eqref{eq:H_eff_floquet}.
			(c) Counterclockwise circulation 
			($\varphi = -\pi/3$).
			Parameters are $\omega/J = 40$, 
			$A/\omega \approx 2.37$. The initial state 
			is $|1\rangle$. Time is in units of $1/J$.
		}
	\end{figure}

	\section{Floquet engineering}
	We now demonstrate that perfect chiral circulation can be realized in an open-boundary three-site chain with periodically driven on-site potentials [Fig.~\ref{fig:floquet}(a)]. The driven Hamiltonian reads
	\begin{equation}
		\hat{H}(t) = -J \sum_{j=1}^{2} (\hat{a}^\dagger_{j+1} \hat{a}_j + \text{H.c.}) + \sum_{j=1,3} f_j(t)\, \hat{n}_j,
		\label{eq:H_driven}
	\end{equation}
	with $f_1(t) = A\cos(\omega t + \varphi)$ and $f_3(t) = -A\cos(\omega t - \varphi)$. Here $J$ is the bare NN hopping amplitude, $\hat{n}_j = \hat{a}^\dagger_j \hat{a}_j$ is the number operator, $A$ is the drive amplitude, $\omega$ the drive frequency, and $\varphi$ the drive phase. In the high-frequency limit ($J \ll \omega$)~\cite{eckardt2017colloquium, bukov2015universal}, the effective Hamiltonian obtained from the Magnus expansion reads (see Appendix~\ref{sec:S5}):
	\begin{equation}
		\hat{H}_{\text{eff}} = -J_{\text{eff}} (\hat{a}^\dagger_2 \hat{a}_1 + \hat{a}^\dagger_3 \hat{a}_2) + \tilde{J}\, \hat{a}^\dagger_3 \hat{a}_1 + \text{H.c.},
		\label{eq:H_eff_floquet}
	\end{equation}
	where $J_{\text{eff}} = J\, \mathcal{J}_0(A/\omega)$ is the renormalized NN hopping, $\mathcal{J}_m$ denotes the $m$th-order Bessel function of the first kind, and
	\begin{equation}
		\tilde{J} = -i\frac{2J^2}{\omega}\sum_{m=1}^{\infty}\frac{1}{m}\,\mathcal{J}_m(A/\omega)\,\mathcal{J}_{-m}(A/\omega)\,\sin(2m\varphi)
		\label{eq:J_tilde}
	\end{equation}
	is the induced NNN coupling arising from the first-order Magnus correction. The commutator structure dictates a purely imaginary prefactor for $\tilde{J}$ in Eq.~\eqref{eq:H_eff_floquet}, fixing the NNN hopping phase to $\pm\pi/2$ and the total flux to $\Phi = \pm\pi/2$. Diagonalizing this Hamiltonian yields the characteristic equation $E^3 - (2J_{\text{eff}}^2 + |\tilde{J}|^2)E = 0$, whose symmetric and equally spaced eigenvalues inherently satisfy the equidistant spectrum requirement. However, realizing perfect chiral circulation also requires restoring discrete translational invariance. This is achieved by enforcing equal bond strengths, $|\tilde{J}| = |J_{\text{eff}}|$, which can be satisfied by tuning parameters (e.g., $A/\omega \approx 2.37$ for $\omega/J = 40$ and $\varphi = \pi/3$). Under this condition, a gauge transformation distributes the total flux equally among the three bonds as uniform chiral phases. This fulfills the translational invariance requirement, establishing the exact target Hamiltonian. Full time-dependent simulations [Fig.~\ref{fig:floquet}(b),(c)] confirm excellent agreement with this effective model.
	
	\begin{figure}[tb]
		\centering
		\includegraphics[width=0.9\linewidth]{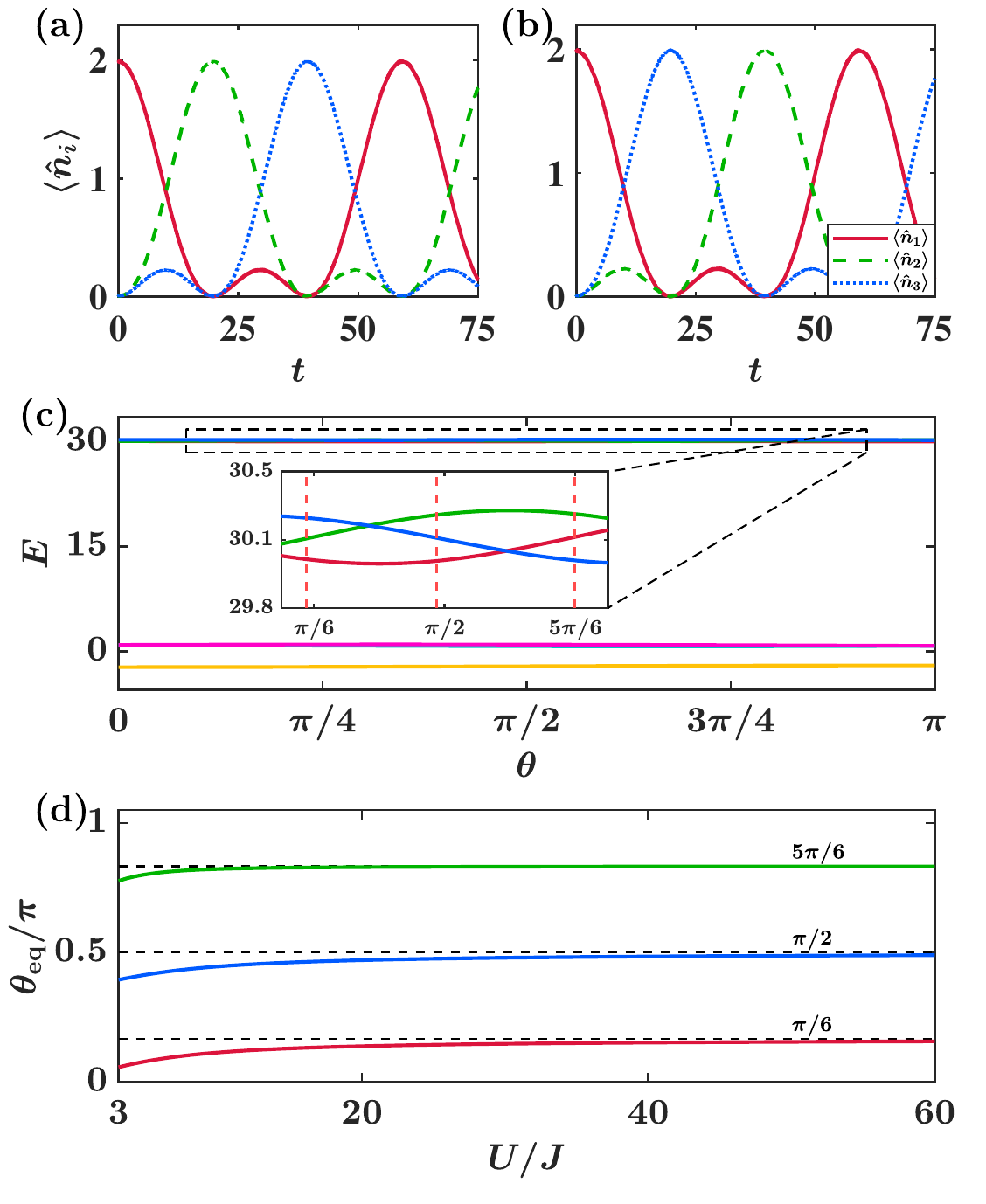}
		\caption{\label{fig:anyon}  
			Chiral transport via anyonic statistics. 
			(a) Clockwise doublon circulation for a statistical phase $\theta = \pi/6$. Lines show the particle number expectations $\langle \hat{n}_i \rangle$ for sites $i=1$ (red), $2$ (green), and $3$ (blue), starting with two particles at site~1. 
			(b) Counterclockwise circulation for $\theta = -\pi/6$. 
			(c) Full two-particle energy spectrum (six levels) versus $\theta$. Inset: the three doublon levels, where the dashed lines mark the values of $\theta_{\text{eq}}$ at which they become equidistant. 
			(d) Equidistant angle $\theta_{\text{eq}}$ versus $U/J$. As $U$ increases, the three branches $\theta_{\text{eq}}$ converge to $\pi/6$, $\pi/2$, and $5\pi/6$, as predicted by the effective doublon Hamiltonian~\eqref{eq:H_eff_anyon}, all yielding $|\Phi_{\text{eff}}| = \pi/2$ 
			(mod $2\pi$) and hence perfect chiral circulation. Parameters: $U/J = 30$ in (a)--(c), and time is in units of $1/J$.
		}
	\end{figure}
	
	\section{Anyonic dynamics}
	The Anyon-Hubbard
	model~\cite{keilmann2011statistically,
		greschner2015anyon} provides an alternative
	route in which the fractional statistics
	intrinsically induces the requisite chiral flux.
	The model can be mapped onto a Bose-Hubbard model with density-dependent Peierls phases:
	\begin{equation}
		\hat{H} = -J \sum_{j=1}^{3} \left( 
		e^{-i \theta \hat{n}_j} 
		\hat{b}^\dagger_{j+1} \hat{b}_j 
		+ \text{H.c.} \right) 
		+ \frac{U}{2} \sum_{j=1}^{3} 
		\hat{n}_j(\hat{n}_j-1),
		\label{eq:Anyon_H}
	\end{equation}
	where $\hat{b}_j$ are bosonic operators, $\hat{n}_j = \hat{b}^\dagger_j \hat{b}_j$, $\theta$ is the statistical angle, $U$ the on-site interaction strength, and periodic boundary conditions are imposed via $\hat{b}_{4} \equiv \hat{b}_{1}$. Here we consider two particles on the three-site ring. In the strong-interaction regime $U \gg J$, two particles occupying the same site form a doublon. Within the degenerate subspace spanned by $\{|2\rangle_j\}$ (where $|2\rangle_j$ denotes two particles at site $j$), we define the effective creation operator via $|2\rangle_j = \hat{c}^\dagger_j|0\rangle$ and apply second-order perturbation theory to yield the effective doublon Hamiltonian (see Appendix~\ref{sec:S6}):
	\begin{equation}
		\hat{H}_{\text{d}} = \frac{2J^2}{U} \sum_{j=1}^{3} e^{-i\theta}\, \hat{c}^\dagger_{j+1} \hat{c}_j + \text{H.c.},
		\label{eq:H_eff_anyon}
	\end{equation}
	where $\hat{c}^\dagger_j$ ($\hat{c}_j$) creates (annihilates) a doublon at site $j$, $\hat{c}_{4} \equiv \hat{c}_{1}$, and a uniform on-site energy has been absorbed into the energy zero point. The accumulated flux over the 
	three-site ring is 
	$\Phi_{\text{eff}} = -3\theta$. Setting 
	$\theta = \pi/6$ yields 
	$\Phi_{\text{eff}} = -\pi/2$, automatically
	satisfying the equidistant-spectrum condition and enabling
	perfect clockwise doublon circulation 
	[Fig.~\ref{fig:anyon}(a)]. Reversing the sign, 
	$\theta = -\pi/6$, yields 
	$\Phi_{\text{eff}} = \pi/2$ and counterclockwise 
	circulation [Fig.~\ref{fig:anyon}(b)].
	
	We define $\theta_{\text{eq}}$ as the statistical angle making the three highest levels equidistant. The full two-particle spectrum for $N=3$ [Fig.~\ref{fig:anyon}(c)] comprises three low-lying scattering states and three high-energy doublon levels (inset). At large $U/J$, these bands separate: the high-energy band forms well-defined doublon bound states, while the low-lying band represents unpaired states. As $\theta$ varies, three distinct $\theta_{\text{eq}}$ values render the doublon band equidistant, approaching $\pi/6$, $\pi/2$, and $5\pi/6$ at large $U/J$ [Fig.~\ref{fig:anyon}(d)]. All three yield $|\Phi_{\text{eff}}| = \pi/2$ (mod $2\pi$), satisfying the circulation condition, with the sign of $\theta$ dictating the direction. At $U/J \gtrsim 20$, these $\theta_{\text{eq}}$ branches converge to the analytical limits, validating the effective doublon Hamiltonian~\eqref{eq:H_eff_anyon} as a single-particle picture of perfect chiral circulation. Within this picture, its inherent translational invariance and statistical-phase-induced flux—yielding an equidistant spectrum—naturally fulfill our two criteria, demonstrating their universality even in this few-particle regime.

	\section{Discussion and outlook}
	The two physical criteria established in this work---discrete translational invariance and an equidistant energy spectrum---provide a unified framework for perfect chiral circulation. The target Hamiltonian can be directly realized in superconducting qubit arrays~\cite{Song2019}, photonic synthetic frequency dimensions~\cite{Yuan2018PRB}, and classical electrical circuits. For experimental implementation, since the individual hopping phases are gauge-dependent, the physically relevant quantity is the total magnetic flux through each closed loop. Notably, a gauge transformation exists that sets every pairwise coupling phase to $\pm\pi/2$, ensuring that the enclosed flux per triangle equals $\mp\pi/2$ (see Appendix~\ref{sec:S2}). Because of this gauge freedom, implementing such $\pm\pi/2$ coupling phases is readily achievable in superconducting qubit networks~\cite{Roushan2017} and classical electrical circuits~\cite{Zhang2021PRL,Lin2025APL}.
	
	For the minimal $N=3$ ring, we demonstrate two physically distinct realizations---Floquet engineering and correlated anyonic dynamics---both of which yield the same target Hamiltonian, illustrating the platform-independent nature of our criteria. Both schemes are readily accessible in current cold-atom experiments. In the Floquet scheme, the bare hopping $J \sim \text{kHz}$ sets the time scale to milliseconds, with driving frequencies reaching $\omega/J \sim 10$--$10^2$  achievable in shaken optical lattices~\cite{PhysRevLett.99.220403,PhysRevA.79.013611, PhysRevLett.102.100403}. For the anyonic route, the anyon-Hubbard model can be realized using density-dependent Peierls phases~\cite{keilmann2011statistically,Gorg2019}, where the adjustable statistical angle $\theta$ and a large interaction-to-hopping ratio of $U/J \gtrsim 20$ are experimentally accessible.
	
	Beyond fundamental interest, the chiral-circulation Hamiltonian serves as a deterministic quantum router: the sign of the flux dictates the circulation direction, routing a quantum state from site~$1$ to site~$n+1$ with unit fidelity in a time $n\tilde{T}$. Looking ahead, it is compelling to extend this criterion beyond the linear Hermitian framework to non-Hermitian and nonlinear systems, as well as many-particle and topologically protected chiral transport.
	
	\begin{acknowledgments}
		The work was supported by the National Natural Science Foundation of China (Grants No. 12375022, No. 11975110), and the Zhejiang Sci-Tech University Scientific Research Start-up Fund (Grant No. 20062318-Y).
	\end{acknowledgments}
	
	\bibliographystyle{apsrev4-2}
	\bibliography{chugao1}

@article{cooper2019topological,
	title   = {Topological bands for ultracold atoms},
	author  = {Cooper, N. R. and Dalibard, J. and Spielman, I. B.},
	journal = {Rev. Mod. Phys.},
	volume  = {91},
	pages   = {015005},
	year    = {2019},
	doi     = {10.1103/RevModPhys.91.015005},
}

@article{goldman2016topological,
	title   = {Topological quantum matter with ultracold gases in optical lattices},
	author  = {Goldman, Nathan and Budich, Jan C. and Zoller, Peter},
	journal = {Nat. Phys.},
	volume  = {12},
	pages   = {639--645},
	year    = {2016},
	doi     = {10.1038/nphys3803},
}

@article{Ozawa2019,
	title   = {Topological photonics},
	author  = {Ozawa, Tomoki and Price, Hannah M. and Amo, Alberto and Goldman, Nathan and Hafezi, Mohammad and Lu, Ling and Rechtsman, Mikael C. and Schuster, David and Simon, Jonathan and Zilberberg, Oded and Carusotto, Iacopo},
	journal = {Rev. Mod. Phys.},
	volume  = {91},
	pages   = {015006},
	year    = {2019},
	doi     = {10.1103/RevModPhys.91.015006},
}

@article{Blais2021,
	title   = {Circuit quantum electrodynamics},
	author  = {Blais, Alexandre and Grimsmo, Arne L. and Girvin, S. M. and Wallraff, Andreas},
	journal = {Rev. Mod. Phys.},
	volume  = {93},
	pages   = {025005},
	year    = {2021},
	doi     = {10.1103/RevModPhys.93.025005},
}

@article{PhysRevLett.84.2358,
	title   = {Directed Current due to Broken Time-Space Symmetry},
	author  = {Flach, S. and Yevtushenko, O. and Zolotaryuk, Y.},
	journal = {Phys. Rev. Lett.},
	volume  = {84},
	pages   = {2358--2361},
	year    = {2000},
	doi     = {10.1103/PhysRevLett.84.2358},
}

@article{Denisov2002,
	title   = {Broken space-time symmetries and mechanisms of rectification of ac fields by nonlinear (non)adiabatic response},
	author  = {Denisov, S. and Flach, S. and Ovchinnikov, A. A. and Yevtushenko, O. and Zolotaryuk, Y.},
	journal = {Phys. Rev. E},
	volume  = {66},
	pages   = {041104},
	year    = {2002},
	doi     = {10.1103/PhysRevE.66.041104},
}

@article{RevModPhys.81.387,
	title   = {Artificial {B}rownian motors: Controlling transport on the nanoscale},
	author  = {H{\"a}nggi, Peter and Marchesoni, Fabio},
	journal = {Rev. Mod. Phys.},
	volume  = {81},
	pages   = {387--442},
	year    = {2009},
	doi     = {10.1103/RevModPhys.81.387},
}

@article{reimann2002brownian,
	title   = {Brownian motors: noisy transport far from equilibrium},
	author  = {Reimann, Peter},
	journal = {Phys. Rep.},
	volume  = {361},
	pages   = {57--265},
	year    = {2002},
	doi     = {10.1016/S0370-1573(01)00081-3},
}

@article{PhysRevLett.131.133401,
	title   = {Hamiltonian Ratchet for Matter-Wave Transport},
	author  = {Dupont, N. and Gabardos, L. and Arrouas, F. and Ombredane, N. and Billy, J. and Peaudecerf, B. and Gu{\'e}ry-Odelin, D.},
	journal = {Phys. Rev. Lett.},
	volume  = {131},
	pages   = {133401},
	year    = {2023},
	doi     = {10.1103/PhysRevLett.131.133401},
}

@article{PhysRevB.27.6083,
	title   = {Quantization of particle transport},
	author  = {Thouless, D. J.},
	journal = {Phys. Rev. B},
	volume  = {27},
	pages   = {6083--6087},
	year    = {1983},
	doi     = {10.1103/PhysRevB.27.6083},
}

@article{Citro2023,
	title   = {Thouless pumping and topology},
	author  = {Citro, Roberta and Aidelsburger, Monika},
	journal = {Nat. Rev. Phys.},
	volume  = {5},
	pages   = {87--101},
	year    = {2023},
	doi     = {10.1038/s42254-022-00545-0},
}

@article{TKNN1982,
	title     = {Quantized {Hall} Conductance in a Two-Dimensional Periodic Potential},
	author    = {Thouless, D. J. and Kohmoto, M. and Nightingale, M. P. and den Nijs, M.},
	journal   = {Phys. Rev. Lett.},
	volume    = {49},
	pages     = {405--408},
	year      = {1982},
	doi       = {10.1103/PhysRevLett.49.405},
	publisher = {American Physical Society},
}

@article{FuYe2022PRL,
	title   = {Nonlinear Thouless Pumping: Solitons and Transport Breakdown},
	author  = {Fu, Qidong and Wang, Peng and Kartashov, Yaroslav V. and Konotop, Vladimir V. and Ye, Fangwei},
	journal = {Phys. Rev. Lett.},
	volume  = {128},
	pages   = {154101},
	year    = {2022},
	doi     = {10.1103/PhysRevLett.128.154101},
}

@article{HuXX2024,
	title   = {Pumping of matter wave solitons in one-dimensional optical superlattices},
	author  = {Hu, Xiaoxiao and Li, Zhiqiang and Chen, Ai-Xi and Luo, Xiaobing},
	journal = {New J. Phys.},
	volume  = {26},
	pages   = {123006},
	year    = {2024},
	doi     = {10.1088/1367-2630/ad9770},
}

@article{Roushan2017,
	title   = {Chiral ground-state currents of interacting photons in a synthetic magnetic field},
	author  = {Roushan, P. and Neill, C. and Megrant, A. and Chen, Y. and Babbush, R. and Barends, R. and Campbell, B. and Chen, Z. and Chiaro, B. and Dunsworth, A. and Fowler, A. and Jeffrey, E. and Kelly, J. and Lucero, E. and Mutus, J. and O'Malley, P. J. J. and Neeley, M. and Quintana, C. and Sank, D. and Vainsencher, A. and Wenner, J. and White, T. and Kapit, E. and Neven, H. and Martinis, J.},
	journal = {Nat. Phys.},
	volume  = {13},
	pages   = {146--151},
	year    = {2017},
	doi     = {10.1038/nphys3930},
}

@article{Koch2010,
	title   = {Time-reversal-symmetry breaking in circuit-{QED}-based photon lattices},
	author  = {Koch, Jens and Houck, Andrew A. and Le Hur, Karyn and Girvin, S. M.},
	journal = {Phys. Rev. A},
	volume  = {82},
	pages   = {043811},
	year    = {2010},
	doi     = {10.1103/PhysRevA.82.043811},
}

@article{Wang2019,
	title   = {Synthesis of antisymmetric spin exchange interaction and chiral spin clusters in superconducting circuits},
	author  = {Wang, Da-Wei and Song, Chao and Feng, Wei and Cai, Han and Xu, Da and Deng, Hui and Li, Hekang and Zheng, Dongning and Zhu, Xiaobo and Wang, H. and Zhu, Shi-Yao and Scully, Marlan O.},
	journal = {Nat. Phys.},
	volume  = {15},
	pages   = {382--386},
	year    = {2019},
	doi     = {10.1038/s41567-018-0400-9},
}

@article{Downing2020,
	title   = {Chiral Current Circulation and {PT} Symmetry in a Trimer of Oscillators},
	author  = {Downing, Charles A. and Zueco, David and Mart{\'i}n-Moreno, Luis},
	journal = {ACS Photonics},
	volume  = {7},
	pages   = {3401--3414},
	year    = {2020},
	doi     = {10.1021/acsphotonics.0c01208},
}

@article{Herrmann2022,
	title   = {Mirror symmetric on-chip frequency circulation of light},
	author  = {Herrmann, Jason F. and Ansari, Vahid and Wang, Jiahui and Witmer, Jeremy D. and Fan, Shanhui and Safavi-Naeini, Amir H.},
	journal = {Nat. Photonics},
	volume  = {16},
	pages   = {603--608},
	year    = {2022},
	doi     = {10.1038/s41566-022-01026-7},
}

@article{PhysRevLett.126.123603,
	title   = {Synthetic Gauge Fields in a Single Optomechanical Resonator},
	author  = {Chen, Yuan and Zhang, Yan-Lei and Shen, Zhen and Zou, Chang-Ling and Guo, Guang-Can and Dong, Chun-Hua},
	journal = {Phys. Rev. Lett.},
	volume  = {126},
	pages   = {123603},
	year    = {2021},
	doi     = {10.1103/PhysRevLett.126.123603},
}

@article{Qi2022,
	title   = {Chiral current in {F}loquet cavity-magnonics},
	author  = {Qi, Shi-fan and Jing, Jun},
	journal = {Phys. Rev. A},
	volume  = {106},
	pages   = {033711},
	year    = {2022},
	doi     = {10.1103/PhysRevA.106.033711},
}

@article{PhysRevApplied.23.054080,
	title   = {Chiral excitation flows of a multinode network based on synthetic gauge fields},
	author  = {Lu, Xian-Liang and Wang, Fo-Hong and Zou, Jia-Jin and Xiang, Ze-Liang},
	journal = {Phys. Rev. Appl.},
	volume  = {23},
	pages   = {054080},
	year    = {2025},
	doi     = {10.1103/PhysRevApplied.23.054080},
}

@article{bose2003quantum,
	title   = {Quantum Communication through an Unmodulated Spin Chain},
	author  = {Bose, Sougato},
	journal = {Phys. Rev. Lett.},
	volume  = {91},
	pages   = {207901},
	year    = {2003},
	doi     = {10.1103/PhysRevLett.91.207901},
}

@article{christandl2004perfect,
	title   = {Perfect State Transfer in Quantum Spin Networks},
	author  = {Christandl, Matthias and Datta, Nilanjana and Ekert, Artur and Landahl, Andrew J.},
	journal = {Phys. Rev. Lett.},
	volume  = {92},
	pages   = {187902},
	year    = {2004},
	doi     = {10.1103/PhysRevLett.92.187902},
}

@article{Kay2010,
	title   = {Perfect, efficient, state transfer and its application as a constructive tool},
	author  = {Kay, Alastair},
	journal = {Int. J. Quantum Inf.},
	volume  = {8},
	pages   = {641--676},
	year    = {2010},
	doi     = {10.1142/S0219749910006514},
}

@article{eckardt2017colloquium,
	title   = {Colloquium: Atomic quantum gases in periodically driven optical lattices},
	author  = {Eckardt, Andr{\'e}},
	journal = {Rev. Mod. Phys.},
	volume  = {89},
	pages   = {011004},
	year    = {2017},
	doi     = {10.1103/RevModPhys.89.011004},
}

@article{bukov2015universal,
	title   = {Universal high-frequency behavior of periodically driven systems: from dynamical stabilization to {F}loquet engineering},
	author  = {Bukov, Marin and D'Alessio, Luca and Polkovnikov, Anatoli},
	journal = {Adv. Phys.},
	volume  = {64},
	pages   = {139--226},
	year    = {2015},
	doi     = {10.1080/00018732.2015.1055918},
}

@article{keilmann2011statistically,
	title   = {Statistically induced phase transitions and anyons in {1D} optical lattices},
	author  = {Keilmann, Tassilo and Lanzmich, Simon and McCulloch, Ian and Roncaglia, Marco},
	journal = {Nat. Commun.},
	volume  = {2},
	pages   = {361},
	year    = {2011},
	doi     = {10.1038/ncomms1353},
}

@article{greschner2015anyon,
	title   = {Anyon {H}ubbard Model in One-Dimensional Optical Lattices},
	author  = {Greschner, Sebastian and Santos, Luis},
	journal = {Phys. Rev. Lett.},
	volume  = {115},
	pages   = {053002},
	year    = {2015},
	doi     = {10.1103/PhysRevLett.115.053002},
}

@article{Song2019,
	title   = {Generation of multicomponent atomic {S}chr{\"o}dinger cat states of up to 20 qubits},
	author  = {Song, Chao and Xu, Kai and Li, Hekang and Zhang, Yu-Ran and Zhang, Xu and Liu, Wuxin and Guo, Qiujiang and Wang, Zhen and Ren, Wenhui and Hao, Jie and Feng, Hui and Fan, Heng and Zheng, Dongning and Wang, Da-Wei and Wang, H. and Zhu, Shi-Yao},
	journal = {Science},
	volume  = {365},
	pages   = {574--577},
	year    = {2019},
	doi     = {10.1126/science.aay0600},
}

@article{Yuan2018PRB,
	title   = {Synthetic space with arbitrary dimensions in a few rings undergoing dynamic modulation},
	author  = {Yuan, Luqi and Xiao, Meng and Lin, Qian and Fan, Shanhui},
	journal = {Phys. Rev. B},
	volume  = {97},
	pages   = {104105},
	year    = {2018},
	doi     = {10.1103/PhysRevB.97.104105},
}

@article{Zhang2021PRL,
	title   = {Experimental Observation of Higher-Order Topological Anderson Insulators},
	author  = {Zhang, Weixuan and Zou, Deyuan and Pei, Qingsong and He, Wenjing and Bao, Jiacheng and Sun, Houjun and Zhang, Xiangdong},
	journal = {Phys. Rev. Lett.},
	volume  = {126},
	pages   = {146802},
	year    = {2021},
	doi     = {10.1103/PhysRevLett.126.146802},
}

@article{Lin2025APL,
	title   = {Manipulating the non-{H}ermitian skin effect through nonreciprocal flux in topolectrical circuits},
	author  = {Lin, Wei and Liu, Chao and Ruan, Banxian and Zou, Yanhong and Dai, Xiaoyu and Xiang, Yuanjiang},
	journal = {Appl. Phys. Lett.},
	volume  = {127},
	pages   = {133306},
	year    = {2025},
	doi     = {10.1063/5.0290682},
}

@article{PhysRevLett.99.220403,
	title   = {Dynamical Control of Matter-Wave Tunneling in Periodic Potentials},
	author  = {Lignier, H. and Sias, C. and Ciampini, D. and Singh, Y. and Zenesini, A. and Morsch, O. and Arimondo, E.},
	journal = {Phys. Rev. Lett.},
	volume  = {99},
	pages   = {220403},
	year    = {2007},
	doi     = {10.1103/PhysRevLett.99.220403},
}

@article{PhysRevA.79.013611,
	title   = {Exploring dynamic localization with a {B}ose-{E}instein condensate},
	author  = {Eckardt, Andr{\'e} and Holthaus, Martin and Lignier, Hans and Zenesini, Alessandro and Ciampini, Donatella and Morsch, Oliver and Arimondo, Ennio},
	journal = {Phys. Rev. A},
	volume  = {79},
	pages   = {013611},
	year    = {2009},
	doi     = {10.1103/PhysRevA.79.013611},
}

@article{PhysRevLett.102.100403,
	title   = {Coherent Control of Dressed Matter Waves},
	author  = {Zenesini, Alessandro and Lignier, Hans and Ciampini, Donatella and Morsch, Oliver and Arimondo, Ennio},
	journal = {Phys. Rev. Lett.},
	volume  = {102},
	pages   = {100403},
	year    = {2009},
	doi     = {10.1103/PhysRevLett.102.100403},
}

@article{Gorg2019,
	title   = {Realization of density-dependent {P}eierls phases to engineer quantized gauge fields coupled to ultracold matter},
	author  = {G{\"o}rg, Frederik and Sandholzer, Kilian and Minguzzi, Joaqu{\'\i}n and Desbuquois, R{\'e}mi and Messer, Michael and Esslinger, Tilman},
	journal = {Nat. Phys.},
	volume  = {15},
	pages   = {1161--1167},
	year    = {2019},
	doi     = {10.1038/s41567-019-0615-4},
}
	
	\appendix
	\clearpage
	\section{Equidistant Spectrum as a Prerequisite for Chiral Circulation}
	\label{sec:S1}
	
	We analyze the spectral properties enabling a Hamiltonian $\hat{H}$ to generate perfect sequential cyclic state transfer across $N$ sites with basis states $\{|1\rangle, \dots, |N\rangle\}$.
	
	Perfect sequential state transfer with a single-step period $\tilde{T}$ requires the system to evolve from site $|n\rangle$ to $|n+1\rangle$ at each interval $\tilde{T}$. Specifically, the time-evolution operator $\hat{U}(\tilde{T}) = e^{-i\hat{H}\tilde{T}}$ must satisfy:
	\begin{equation}
		\hat{U}(\tilde{T})|n\rangle = e^{i\phi_n}|n+1\rangle \quad \text{for } n=1,\dots,N-1,
		\label{eq:evolution-bulk}
	\end{equation}
	along with the cyclic boundary condition:
	\begin{equation}
		\hat{U}(\tilde{T})|N\rangle = e^{i\phi_N}|1\rangle,
		\label{eq:evolution-cyclic}
	\end{equation}
	where $\phi_n$ is an arbitrary phase factor accumulated during transfer.
	
	The matrix elements of $\hat{U}(\tilde{T})$ are given by $\hat{U}_{mn} = \langle m | \hat{U}(\tilde{T}) | n \rangle$. Conditions~\eqref{eq:evolution-bulk} and~\eqref{eq:evolution-cyclic} dictate that $\hat{U}(\tilde{T})$ is a generalized permutation matrix:
	\begin{equation}
		\hat{U}(\tilde{T}) = \begin{pmatrix}
			0 & 0 & \dots & 0 & e^{i\phi_N} \\
			e^{i\phi_1} & 0 & \dots & 0 & 0 \\
			0 & e^{i\phi_2} & \dots & 0 & 0 \\
			\vdots & \vdots & \ddots & \vdots & \vdots \\
			0 & 0 & \dots & e^{i\phi_{N-1}} & 0
		\end{pmatrix}.
		\label{eq:evolution-matrix}
	\end{equation}
	
	To constrain $\hat{H}$, we evaluate the characteristic polynomial $\det(\hat{U}(\tilde{T}) - \lambda I) = 0$. Expanding along the first row yields only two non-zero terms:
	\begin{equation}
		\det(\hat{U} - \lambda I) = (-\lambda)\det(M_{11}) + e^{i\phi_N}(-1)^{N+1}\det(M_{1N}),
		\label{eq:laplace-expansion}
	\end{equation}
	where $M_{11}$ is a lower triangular matrix with diagonal entries $-\lambda$ [giving $\det(M_{11})=(-\lambda)^{N-1}$], and $M_{1N}$ is an upper triangular matrix with diagonal entries $e^{i\phi_n}$ [giving $\det(M_{1N})=\prod_{n=1}^{N-1}e^{i\phi_n}$]. Substituting these determinants back gives the characteristic equation:
	\begin{equation}
		(-\lambda)^N + (-1)^{N+1} e^{i\Theta} = 0,
		\label{eq:char-equation}
	\end{equation}
	where $\Theta = \sum_{n=1}^N \phi_n$ is the total accumulated phase. The solutions are uniformly distributed on the unit circle:
	\begin{equation}
		\lambda_k = \exp\left( i \frac{\Theta - 2\pi k}{N} \right), \quad k = 0, 1, \dots, N-1.
		\label{eq:U-eigenvalues}
	\end{equation}
	
	Relating these to the energy spectrum of $\hat{H}$ via $\lambda_k = e^{-iE_k\tilde{T}}$, and accounting for the branch ambiguity $m \in \mathbb{Z}$ of the complex logarithm, we obtain:
	\begin{equation}
		E_k \tilde{T} = -\frac{\Theta - 2\pi k}{N} - 2\pi m.
		\label{eq:log-condition}
	\end{equation}
	Different choices of $m$ merely shift the entire spectrum by $2\pi m/\tilde{T}$ without altering the dynamics.  Because the eigenvalues ${\lambda_k}$ are equally spaced on the unit circle with a uniform angular separation of $2\pi/N$, the map $\lambda \mapsto E$ necessarily dictates that the energy eigenvalues ${E_k}$ are also equally spaced. \emph{This equidistant spectrum is thereby enforced by the cyclic permutation structure of $\hat{U}(\tilde{T})$}, rather than being imposed as an additional assumption.
	
	Choosing the principal branch ($m = 0$), the spectrum is:
	\begin{equation}
		E_k = \frac{1}{\tilde{T}}\left(\frac{2\pi k}{N} - \frac{\Theta}{N}\right), \quad k = 0,1,\dots,N-1,
		\label{eq:energy-spectrum}
	\end{equation}
	with a uniform level spacing $|\Delta E| = 2\pi/(N\tilde{T}) \equiv b$.
	
	The total phase $\Theta$ is a gauge degree of freedom that shifts the entire spectrum uniformly without affecting the level spacing. Throughout this work, we set $\Theta = 0$ for simplicity, giving $E_k = bk$ to establish a strictly increasing energy ordering. As shown in Appendix~\ref{sec:S2}, this positive energy slope is precisely the condition that drives the clockwise circulation $|n\rangle \to |n+1\rangle$.
	
	\section{Constructing the Ring Hamiltonian from an Equidistant Spectrum} 
	\label{sec:S2}
	
	In Appendix~\ref{sec:S1}, we showed that perfect sequential circulation \emph{requires} an equidistant energy spectrum. Here we demonstrate that translational invariance combined with an equidistant spectrum is also \emph{sufficient}, and reconstruct the unique Hamiltonian satisfying both conditions.
	
	Consider a translationally invariant Hamiltonian $\hat{H}$ with equidistant spectrum $E_k = bk$ ($k = 0,\dots,N-1$), where $b = 2\pi/(N\tilde{T})$ is the uniform level spacing defined in Appendix~\ref{sec:S1}. Since $\hat{H}$ commutes with the discrete translation operator $\hat{T}$, defined by $\hat{T}|n\rangle = |n+1\rangle$ (with periodic boundaries $|N+1\rangle \equiv |1\rangle$), Bloch's theorem guarantees that the eigenstates are the discrete Fourier modes:
	\begin{equation}
		|k\rangle = \frac{1}{\sqrt{N}} \sum_{n=1}^N e^{i \frac{2\pi k n}{N}} |n\rangle, \quad k=0,\dots,N-1,
		\label{eq:bloch-modes}
	\end{equation}
	where $k$ labels the discrete quasi-momentum $p_k = 2\pi k/N$. These modes intrinsically satisfy the Bloch condition $\langle n+1|k\rangle = e^{i\frac{2\pi k}{N}}\langle n|k\rangle$, ensuring compatibility with the periodic boundary conditions.
	
	Applying the inverse Fourier transform, the site states are
	\begin{equation}
		|n\rangle = \frac{1}{\sqrt{N}}\sum_{k=0}^{N-1} e^{-i\frac{2\pi kn}{N}}|k\rangle.
	\end{equation}
	Acting with $\hat{U}(\tilde{T}) = e^{-i\hat{H}\tilde{T}}$ and substituting $E_k = bk$:
	\begin{equation}
		\begin{split}
			\hat{U}(\tilde{T})|n\rangle 
			&= \frac{1}{\sqrt{N}} \sum_{k=0}^{N-1} e^{-i\frac{2\pi kn}{N}} e^{-iE_k \tilde{T}} |k\rangle \\
			&= \frac{1}{\sqrt{N}} \sum_{k=0}^{N-1} e^{-i\frac{2\pi k(n+1)}{N}} |k\rangle = |n+1\rangle,
		\end{split}
		\label{eq:U-circulates-fwd}
	\end{equation}
	where we used $E_k \tilde{T} = bk \tilde{T} = 2\pi k/N$, and the last equality follows from the inverse Fourier transform (applying the cyclic condition $|N+1\rangle \equiv |1\rangle$ when $n=N$). Eq.~\eqref{eq:U-circulates-fwd} confirms that translational invariance together with the equidistant spectrum $E_k = bk$ ($b > 0$) is sufficient to produce perfect clockwise circulation $|n\rangle \to |n+1\rangle$ with unit fidelity. Combined with the necessary condition derived in Appendix~\ref{sec:S1}, we arrive at the following conclusion:
	
	\medskip
	\noindent\textit{For a translationally invariant Hamiltonian, an equidistant energy spectrum is the necessary and sufficient condition for perfect sequential cyclic state transfer $|1\rangle\to|2\rangle\to\cdots\to|N\rangle\to|1\rangle$ with unit fidelity.}
	\medskip
	
	We now reconstruct the unique translationally invariant Hamiltonian with equidistant spectrum $E_k = bk$. Since the Hamiltonian matrix elements depend only on the relative distance $m - n$, they are obtained in the site basis as:
	\begin{equation}
		\begin{split}
			\hat{H}_{mn} &= \sum_{k} E_k \langle m | k \rangle \langle k | n \rangle \\
			&= \sum_{k=0}^{N-1} (bk) \cdot \frac{1}{\sqrt{N}} e^{i\frac{2\pi km}{N}} \cdot \frac{1}{\sqrt{N}} e^{-i\frac{2\pi kn}{N}} \\
			&= \frac{b}{N} \sum_{k=0}^{N-1} k\, e^{i \frac{2\pi k (m-n)}{N}}.
		\end{split}
		\label{eq:H-fourier}
	\end{equation}
	For the diagonal elements ($m = n$), this yields $\hat{H}_{mm} = b(N-1)/2$, which is a uniform on-site energy independent of $m$. This constant does not affect the dynamics and is absorbed into the energy zero point hereafter.
	
	For the off-diagonal elements ($m \neq n$), we apply the identity $\sum_{k=0}^{N-1} k\, x^k = N/(x-1)$ with $x = e^{i 2\pi (m-n)/N}$, yielding
	\begin{equation}
		\hat{H}_{mn} = \frac{b}{e^{i \frac{2\pi (m-n)}{N}} - 1}.
		\label{eq:H-hopping}
	\end{equation}
	Rewriting this in polar form, we obtain
	\begin{equation}
		\hat{H}_{mn} = \frac{b}{2\left|\sin\frac{\pi(m-n)}{N}\right|} 
		\, e^{-i\left(\frac{\pi(m-n)}{N} 
			+ \frac{\pi}{2}\,\text{sgn}(m-n)\right)},
		\label{eq:H-hopping-final}
	\end{equation}
	where $\text{sgn}(m-n)$ is the sign function, taking $+1$ for $m > n$ and $-1$ for $m < n$.
	
	The Hamiltonian can be written in second-quantized form as
	\begin{equation}
		\hat{H} = -\sum_{m > n} \left( |J_{mn}|\, e^{i\Phi_{mn}}\, \hat{a}^\dagger_m \hat{a}_n + \text{H.c.} \right),
		\label{eq:H-ideal}
	\end{equation}
	with hopping amplitude and phase:
	\begin{subequations}
		\begin{align}
			|J_{mn}| &= \frac{b}{2 \left|\sin\left(\frac{\pi}{N}(m-n)\right)\right|}, \label{eq:hopping-amplitude} \\
			\Phi_{mn} &= -\frac{\pi}{N}(m-n) + \frac{\pi}{2}, \quad (m > n). \label{eq:hopping-phase}
		\end{align}
	\end{subequations}
	
	Since the total flux piercing the ring is gauge invariant, one can always perform a local phase shift $\Phi_{mn} \to \Phi_{mn} + \alpha_m - \alpha_n$ without changing the physics. This is because the accumulation of the additional phase $\alpha_m - \alpha_n$ along any closed loop forms a telescoping sum that cancels to zero. Utilizing this freedom, a gauge transformation with $\alpha_n = \pi n/N$ shifts every bond phase in Eq.~\eqref{eq:hopping-phase} by $\Phi_{mn} \to \Phi_{mn} + \frac{\pi(m-n)}{N}$. This transformation corresponds to the diagonal unitary $\hat{D} = \mathrm{diag}(e^{i\alpha_1}, \ldots, e^{i\alpha_N})$, which redefines the site basis via $\hat{D}|n\rangle = e^{i\alpha_n}|n\rangle$ and transforms the matrix elements as $\hat{H}_{mn} \to e^{i\frac{\pi(m-n)}{N}} \hat{H}_{mn}$ under the similarity transformation $\hat{H} \to \hat{D}\,\hat{H}\,\hat{D}^\dagger$. Consequently, the coupling phase is rendered uniform, yielding $\Phi_{mn} = \pi/2$ for all bonds.
	
	\section{Reversing the Direction of Circulation}
	\label{sec:S3}
	
	In Appendix~\ref{sec:S1}, the clockwise circulation $|n\rangle \to |n+1\rangle$ corresponds to the evolution operator $\hat{U}(\tilde{T}) = e^{-i\hat{H}\tilde{T}}$ and the equidistant spectrum $E_k = bk$. We now consider the counterclockwise circulation and derive the Hamiltonian that governs this reverse dynamics.
	
	Define the site-reversal matrix:
	\begin{equation}
		\hat{P} = \begin{pmatrix}
			0 & 0 & \dots & 0 & 1 \\
			0 & 0 & \dots & 1 & 0 \\
			\vdots & & \iddots & & \vdots \\
			0 & 1 & \dots & 0 & 0 \\
			1 & 0 & \dots & 0 & 0
		\end{pmatrix},
	\end{equation}
	which reverses the site indices:
	\begin{equation}
		\hat{P}|n\rangle = |N+1-n\rangle.
	\end{equation}
	It is straightforward to verify that $\hat{P}^\dagger \hat{P} = \hat{P}^2 = I$, which implies that $\hat{P} = \hat{P}^{-1} = \hat{P}^\dagger$.
	
	By definition, the clockwise evolution operator acts as $\hat{U}(\tilde{T})|n\rangle = e^{i\phi_n}|n+1\rangle$. To find the operator for backward circulation, we define $\hat{U}'(\tilde{T}) \equiv \hat{P}\hat{U}(\tilde{T})\hat{P}$. Applying this to a state $|n\rangle$ and utilizing $\hat{P}^2 = I$, we obtain
	\begin{equation}
		\begin{split}
			\hat{U}'(\tilde{T})|n\rangle &= \hat{P}\hat{U}(\tilde{T})\hat{P}|n\rangle \\
			&= \hat{P}\hat{U}(\tilde{T})|N+1-n\rangle \\
			&= e^{i\phi_{N+1-n}}\hat{P}|N+2-n\rangle \\
			&= e^{i\phi_{N+1-n}}|n-1\rangle,
		\end{split}
	\end{equation}
	which confirms that $\hat{U}'(\tilde{T})$ realizes the counterclockwise circulation $|n\rangle \to |n-1\rangle$.
	
	Since $\hat{U}'(\tilde{T}) = e^{-i\hat{H}'\tilde{T}}$, we identify the corresponding Hamiltonian as $\hat{H}' = \hat{P}\hat{H}\hat{P}$. This unitary similarity transformation guarantees that $\hat{H}'$ and $\hat{H}$ share the same energy spectrum.
	
	The matrix elements of the Hamiltonian $\hat{H}'$ for perfect counterclockwise circulation take the form
	\begin{equation}
		\begin{split}
			\hat{H}'_{mn} &= \langle m|\hat{P}\hat{H}\hat{P}|n\rangle \\
			&= \langle N+1-m|\hat{H}|N+1-n\rangle \\
			&= \hat{H}_{(N+1-m)(N+1-n)}.
		\end{split}
		\label{eq:Hprime-matrix}
	\end{equation}
	Discrete translational invariance implies $(\hat{T}^\dagger)^l \hat{H}\, \hat{T}^l = \hat{H}$ for any integer $l$. To connect this to $\hat{H}_{nm}$, we note that translating the site $N+1-n$ forward by $l = m+n-N-1$ steps shifts it precisely to site $m$. Applying this translation yields
	\begin{equation}
		\begin{split}
			\hat{H}_{(N+1-m)(N+1-n)} 
			&= \langle N+1-m|(\hat{T}^\dagger)^{l} \hat{H}\, \hat{T}^{l}|N+1-n\rangle \\
			&= \langle n|\hat{H}|m\rangle \\
			&= \hat{H}_{nm} = \hat{H}_{mn}^*,
		\end{split}
		\label{eq:Hprime-translate}
	\end{equation}
	where the last equality follows from Hermiticity. Combining Eqs.~\eqref{eq:Hprime-matrix} and \eqref{eq:Hprime-translate}, we obtain
	\begin{equation}
		\hat{H}'_{mn} = \hat{H}_{(N+1-m)(N+1-n)} = \hat{H}_{mn}^*.
	\end{equation}
	
	That is, $\hat{H}' = \hat{H}^*$: the hopping amplitudes $|J_{mn}|$ remain unchanged, while all hopping phases reverse sign ($\Phi_{mn} \to -\Phi_{mn}$), naturally realizing the counterclockwise circulation $|n\rangle \to |n-1\rangle$.
	
	From Appendix~\ref{sec:S2}, the eigenstates of $\hat{H}$ are the Bloch modes
	\begin{equation}
		|k\rangle = \frac{1}{\sqrt{N}} \sum_{n=1}^{N} e^{i\frac{2\pi kn}{N}} |n\rangle, \quad \hat{H}|k\rangle = E_k|k\rangle.
		\label{eq:bloch-forward}
	\end{equation}
	Since $\hat{H}' = \hat{H}^*$, taking the complex conjugate of the eigenvalue equation yields
	\begin{equation}
		\hat{H}'|k\rangle^* = \hat{H}^*|k\rangle^* = E_k|k\rangle^*,
		\label{eq:Hprime-eigen}
	\end{equation}
	where the complex conjugate state is
	\begin{equation}
		\begin{split}
			|k\rangle^* &= \frac{1}{\sqrt{N}} \sum_{n=1}^{N} e^{-i\frac{2\pi kn}{N}} |n\rangle \\
			&= \frac{1}{\sqrt{N}} \sum_{n=1}^{N} e^{i\frac{2\pi (N-k)n}{N}} |n\rangle \\
			&= |N-k\rangle .
		\end{split}
		\label{eq:kstar}
	\end{equation}
	Equivalently, this shows that the state $|k\rangle$ is an eigenstate of $\hat{H}'$ with eigenvalue $E_{N-k}$:
	\begin{equation}
		\hat{H}' |k\rangle = E_{N-k} |k\rangle = b(N-k) |k\rangle.
	\end{equation}
	While $\{E_k\}$ and $\{E_{N-k}\}$ constitute the same set of eigenvalues, their ordering with respect to the quasi-momentum $p_k = 2\pi k/N$ is strictly reversed. Since the group velocity is given by $v_g = \partial E_k / \partial p_k = b N / (2\pi)$, this momentum reversal directly implies a sign change in the group velocity, $v_g \to -v_g$, thereby realizing the counterclockwise circulation.
	
	\section{Numerical Evaluation of Hopping Parameters}
	\label{sec:S4}
	
	To illustrate the hopping parameters defined in Eqs.~\eqref{eq:hopping-amplitude} and~\eqref{eq:hopping-phase}, explicit values for $N=3, 4, 5$ are listed in Table~\ref{tab:params}. Here, $|J_\text{NN}|e^{i\Phi_\text{NN}}$ denotes the nearest-neighbor (NN) coupling, and $|J_\text{NNN}|e^{i\Phi_\text{NNN}}$ the next-nearest-neighbor (NNN) coupling. The sign conventions for $\Phi_\text{NN}$ and $\Phi_\text{NNN}$ follow the arrow directions in Fig.~\ref{fig:scalable}(a) of the main text, where red solid arrows denote NN couplings and blue dashed arrows denote NNN couplings.
	
	\begin{table}[h]
		\centering
		\caption{Hopping parameters for $N=3,4,5$.}
		\label{tab:params}
		\begin{ruledtabular}
			\begin{tabular}{cccc}
				$N$ & $|J_\text{NN}|/|J_\text{NNN}|$ & $\Phi_\text{NN}$ & $\Phi_\text{NNN}$ \\
				\hline
				3\footnotemark[1] & $-$ & $\pi/6$ & $-$ \\
				\hline
				4 & $\sqrt{2}$ & $\pi/4$ & $0$ \\
				\hline
				5 & $\frac{1+\sqrt{5}}{2}$ & $3\pi/10$ & $\pi/10$ \\
			\end{tabular}
		\end{ruledtabular}
		\footnotetext[1]{For $N=3$, all sites are mutual nearest neighbors, so only NN couplings exist.}
	\end{table}
	
	Several noteworthy features emerge from these parameters. As expected, for $N=3$, the absence of NNN couplings reflects the fact that all sites are mutually nearest neighbors with equal coupling strengths. For $N \ge 4$, the introduction of NNN couplings is accompanied by a distinct amplitude ratio: notably, for $N=5$, this ratio equals the golden ratio. Furthermore, the NNN hopping phase vanishes for $N=4$, while for $N=5$, the NN and NNN phases become rationally commensurate ($\Phi_\text{NN} = 3\Phi_\text{NNN} = 3\pi/10$). A unifying and key feature of this coupling structure is that for any three mutually coupled sites, regardless of $N$, the hopping phases accumulated along the closed triangle always sum to $\pm\pi/2$ (mod $2\pi$).
	
	\section{Engineering Chiral Flux via Floquet Driving}
	\label{sec:S5}
	
	In this section, we derive the effective Hamiltonian in the high-frequency limit $J \ll \omega$ using a high-frequency expansion (HFE) in a rotating frame. 
	
	We consider an open three-site model described by $\hat{H}(t) = \hat{H}_{\text{static}} + \hat{V}(t)$. The static part is restricted to nearest-neighbor hopping (with no direct coupling between sites 1 and 3),
	\begin{equation}
		\hat{H}_{\text{static}} = -J \left( \hat{a}^\dagger_2 \hat{a}_1 + \hat{a}^\dagger_3 \hat{a}_2 + \text{H.c.} \right),
		\label{eq:H-static}
	\end{equation}
	and the time-periodic on-site drive is given by $\hat{V}(t) = \sum_{j=1}^{3} f_j(t) \hat{n}_j$, with
	\begin{equation}
		f_j(t) = \begin{cases} A\cos(\omega t + \varphi), & j=1, \\ 0, & j=2, \\ -A\cos(\omega t - \varphi), & j=3. \end{cases}
	\end{equation}
	Sites 1 and 3 are driven with equal amplitude but opposite signs, featuring a relative phase of $2\varphi$, while site 2 remains undriven. This phase-shifted driving protocol breaks time-reversal symmetry and induces a synthetic magnetic flux through the ring.
	
	To avoid calculating an infinite series in the high-frequency expansion, we perform a unitary transformation to a rotating frame,
	\begin{equation}
		\hat{S}(t) = \exp\left[ -i \int_0^t d\tau \hat{V}(\tau) \right] = \exp\left[ -i \sum_j \theta_j(t) \hat{n}_j \right],
	\end{equation}
	where the accumulated phases are
	\begin{equation}
		\theta_j(t) = \int_0^t f_j(\tau) d\tau = \begin{cases} \frac{A}{\omega} \sin(\omega t + \varphi), & j=1, \\ 0, & j=2, \\ -\frac{A}{\omega} \sin(\omega t - \varphi), & j=3. \end{cases}
	\end{equation}
	
	In this rotating frame, the driving potential is absorbed into time-dependent hopping phases, yielding
	\begin{equation}
		\begin{split}
			\hat{\tilde{H}}(t) &= \hat{S}^\dagger(t) \hat{H}(t) \hat{S}(t) - i \hat{S}^\dagger(t) \dot{\hat{S}}(t) \\
			&= -J \left( e^{-i\theta_1} \hat{a}^\dagger_2 \hat{a}_1 + e^{i\theta_3} \hat{a}^\dagger_3 \hat{a}_2 + \text{H.c.} \right).
		\end{split}
		\label{eq:H-rotating}
	\end{equation}
	
	We expand the time-dependent phase factors using the Jacobi-Anger identity $e^{iz \sin(\alpha)} = \sum_{m} \mathcal{J}_m(z) e^{i m \alpha}$, where $\mathcal{J}_m$ is the $m$-th order Bessel function. The Hamiltonian can then be expressed as a Fourier series $\hat{\tilde{H}}(t) = \sum_{m} \hat{H}^{(m)} e^{i m \omega t}$, with the components
	\begin{equation}
		\begin{split}
			\hat{H}^{(m)} =& -J (-1)^m \mathcal{J}_m(A/\omega) \\
			& \times\Big[ e^{im\varphi}\, \hat{a}^\dagger_2 \hat{a}_1 
			+ e^{-im\varphi}\, \hat{a}^\dagger_3 \hat{a}_2 + \text{H.c.} \Big].
		\end{split}
		\label{eq:fourier-components}
	\end{equation}
	Here we have used the Bessel function property $\mathcal{J}_m(-z) = (-1)^m \mathcal{J}_m(z)$ to factor out a common coefficient.
	
	The effective Hamiltonian is given by the Magnus expansion
	\begin{equation}
		\hat{H}_{\text{eff}} = \hat{H}^{(0)} + \sum_{m=1}^{\infty} \frac{1}{m\omega}
		[\hat{H}^{(m)}, \hat{H}^{(-m)}] + \mathcal{O}\left(\frac{1}{\omega^2}\right).
		\label{eq:magnus}
	\end{equation}
	The zeroth-order term is the time-average of the driven Hamiltonian,
	\begin{equation}
		\hat{H}^{(0)} = \frac{1}{T}\int_0^T dt\, \hat{\tilde{H}}(t) = -J \mathcal{J}_0(A/\omega) \left( \hat{a}^\dagger_2 \hat{a}_1 + \hat{a}^\dagger_3 \hat{a}_2 + \text{H.c.} \right),
		\label{eq:H-zeroth}
	\end{equation}
	which represents the original nearest-neighbor hopping renormalized by the zeroth-order Bessel function $\mathcal{J}_0(A/\omega)$.
	
	The first-order correction involves the commutators $[\hat{H}^{(m)}, \hat{H}^{(-m)}]$. Substituting the Fourier components~\eqref{eq:fourier-components} into $\sum_{m=1}^{\infty} \frac{1}{m\omega}[\hat{H}^{(m)}, \hat{H}^{(-m)}]$, we obtain
	\begin{equation}
		\sum_{m=1}^{\infty} \frac{1}{m\omega}[\hat{H}^{(m)}, \hat{H}^{(-m)}] =  \tilde{J}\, \hat{a}^\dagger_3 \hat{a}_1 + \text{H.c.},
	\end{equation}
	with
	\begin{equation}
		\tilde{J} = -i\frac{2J^2}{\omega} \sum_{m=1}^{\infty} \frac{1}{m} \mathcal{J}_m(A/\omega)\, \mathcal{J}_{-m}(A/\omega) \sin(2m\varphi).
	\end{equation}
	The induced next-nearest-neighbor hopping carries a phase of $+\pi/2$ for $\mathrm{Im}(\tilde{J}) > 0$ and $-\pi/2$ for $\mathrm{Im}(\tilde{J}) < 0$. Combining this with the zeroth-order term, the effective triangular Hamiltonian reads
	\begin{equation}
		\hat{H}_{\text{eff}} = -J_{\text{eff}} (\hat{a}^\dagger_2 \hat{a}_1 + \hat{a}^\dagger_3 \hat{a}_2) +  \tilde{J}\, \hat{a}^\dagger_3 \hat{a}_1 + \text{H.c.},
		\label{eq:H-eff-triangular}
	\end{equation}
	where $J_{\text{eff}} = J \mathcal{J}_0(A/\omega)$. 
	
	By tuning the drive frequency $\omega$ and phase $\varphi$, one can satisfy the condition $|J_{\text{eff}}| = |\tilde{J}|$. This synthesizes a three-site Hamiltonian with uniform hopping amplitudes and a total magnetic flux of $\pi/2$ enclosed by the ring, which is precisely the ideal chiral-circulation Hamiltonian required for perfect unidirectional state transfer (see Appendix~\ref{sec:S2}).
	
	\section{Extracting Chiral Flux from Anyonic Bound Pairs}
	\label{sec:S6}
	
	Here we present an alternative route to the ideal chiral-circulation Hamiltonian derived in Appendix~\ref{sec:S1} and~\ref{sec:S2}. In the strong-interaction limit of the Anyon-Hubbard model, the correlated hopping of a two-particle bound state (doublon) naturally generates the requisite chiral phase. We use bosonic operators $\hat{b}_j$---rather than $\hat{a}_j$---to distinguish this interacting model from the single-particle framework of the preceding sections.
	
	The Anyon-Hubbard Hamiltonian, which maps to bosons with density-dependent Peierls phases via the fractional Jordan-Wigner transformation, is decomposed as $\hat{H} = \hat{H}_J + \hat{H}_U$, where
	\begin{equation}
		\hat{H}_J = -J \sum_{j} \left( e^{-i \theta \hat{n}_j} \hat{b}^\dagger_{j+1} \hat{b}_j + \text{H.c.} \right)
	\end{equation}
	represents the hopping term, and
	\begin{equation}
		\hat{H}_U = \frac{U}{2} \sum_{j} \hat{n}_j(\hat{n}_j-1)
	\end{equation}
	denotes the on-site interaction. Here $\hat{n}_j = \hat{b}^\dagger_j \hat{b}_j$ is the particle number operator, $J$ is the hopping amplitude, $U$ is the on-site interaction strength, and $\theta$ is the statistical angle. The phase factor $e^{-i\theta \hat{n}_j}$ encodes the anyonic statistics, distinguishing this model from standard bosons ($\theta=0$).
	
	To derive the effective dynamics of the two-particle bound state (doublon), we define the doublon subspace $\mathcal{P} \equiv \{|2\rangle_j\}$, where $|2\rangle_j$ denotes a state with two particles at site $j$ and zero elsewhere, satisfying $\hat{H}_U |2\rangle_j = E_j |2\rangle_j$ with $E_j = U$; and the separated-particle subspace $\mathcal{Q} \equiv \{|1\rangle_m |1\rangle_l,\, m \neq l\}$, satisfying $\hat{H}_U |1\rangle_m |1\rangle_l = E_{ml} |1\rangle_m |1\rangle_l$ with $E_{ml} = 0$. The corresponding projection operators are
	\begin{equation}
		\hat{\mathcal{P}} = \sum_{j} |2\rangle_j \langle 2|_j, \quad \hat{\mathcal{Q}} = 1 - \hat{\mathcal{P}}.
	\end{equation}
	In the strongly interacting regime $U \gg J$, we treat $\hat{H}_J$ as a perturbation and perform second-order degenerate perturbation theory within the $\mathcal{P}$ subspace. We define the resolvent operator
	\begin{equation}
		\hat{\mathcal{G}} = \sum_{m \neq l} \frac{|1\rangle_m|1\rangle_l \langle 1|_l\langle 1|_m}{U },
	\end{equation}
	where the denominator reflects the energy gap $U$ between the doublon and unpaired states (the two particles occupy distinct sites). The effective Hamiltonian in the $\mathcal{P}$ subspace is given by
	\begin{equation}
		\hat{H}_{\text{eff}} = \hat{h}^{(0)} + \hat{h}^{(1)} + \hat{h}^{(2)} = U\hat{\mathcal{P}} + \hat{\mathcal{P}}\hat{H}_J\hat{\mathcal{P}} + \hat{\mathcal{P}}\hat{H}_J\hat{\mathcal{G}}\hat{H}_J\hat{\mathcal{P}}.
	\end{equation}
	The zero-order term is
	\begin{equation}
		\hat{h}^{(0)} = U\hat{\mathcal{P}} = U\sum_j |2\rangle_j\langle 2|_j.
	\end{equation}
	The first-order term vanishes since single-particle hopping cannot directly connect two doublon states:
	\begin{equation}
		\hat{h}^{(1)} = \hat{\mathcal{P}}\hat{H}_J\hat{\mathcal{P}} = \sum_{j,k} \langle 2|_k\, \hat{H}_J\, |2\rangle_j\, |2\rangle_k\langle 2|_j = 0.
	\end{equation}
	The second-order term describes virtual dissociation and recombination processes:
	\begin{align}
		\hat{h}^{(2)} &= \hat{\mathcal{P}}\hat{H}_J\hat{\mathcal{G}}\hat{H}_J\hat{\mathcal{P}} = \sum_{j,k} h^{(2)}_{kj}\, |2\rangle_k\langle 2|_j,
	\end{align}
	where
	\begin{equation}
		h^{(2)}_{kj} = \sum_{m,l} \frac{\langle 2|_k\, \hat{H}_J\, |1\rangle_m|1\rangle_l\, \langle 1|_l\langle 1|_m\, \hat{H}_J\, |2\rangle_j}{U}.
	\end{equation}
	
	Since $\hat{H}_J$ acting on a doublon $|2\rangle_j$ can only move one particle to an adjacent site, the intermediate state is restricted to $|1\rangle_j|1\rangle_{j+1}$ or $|1\rangle_{j-1}|1\rangle_j$. The three nonvanishing matrix elements are given as follows.
	
	For the forward process ($k=j+1$), the doublon virtually dissociates and recombines via $|2\rangle_j \to |1\rangle_j|1\rangle_{j+1} \to |2\rangle_{j+1}$:
	\begin{align}
		h^{(2)}_{j+1,j} &= \frac{\langle 2|_{j+1}\, \hat{H}_J\, |1\rangle_j|1\rangle_{j+1}\, \langle 1|_{j+1}\langle 1|_j\, \hat{H}_J\, |2\rangle_j}{U} \notag\\
		&= \frac{2J^2}{U} e^{-i\theta}.
	\end{align}
	
	For the backward process ($k=j-1$), the doublon virtually dissociates and recombines via $|2\rangle_j \to |1\rangle_{j-1}|1\rangle_j \to |2\rangle_{j-1}$:
	\begin{align}
		h^{(2)}_{j-1,j} &= \frac{\langle 2|_{j-1}\, \hat{H}_J\, |1\rangle_{j-1}|1\rangle_j\, \langle 1|_j\langle 1|_{j-1}\, \hat{H}_J\, |2\rangle_j}{U} \notag\\
		&= \frac{2J^2}{U} e^{i\theta}.
	\end{align}
	
	For the diagonal process ($k=j$), the doublon virtually dissociates and recombines back at the same site via two channels, $|2\rangle_j \to |1\rangle_j|1\rangle_{j+1} \to |2\rangle_j$ and $|2\rangle_j \to |1\rangle_{j-1}|1\rangle_j \to |2\rangle_j$:
	\begin{align}
		h^{(2)}_{jj} &= \frac{\langle 2|_j\, \hat{H}_J\, |1\rangle_j|1\rangle_{j+1}\, \langle 1|_{j+1}\langle 1|_j\, \hat{H}_J\, |2\rangle_j}{U} \notag\\
		&\quad + \frac{\langle 2|_j\, \hat{H}_J\, |1\rangle_{j-1}|1\rangle_j\, \langle 1|_j\langle 1|_{j-1}\, \hat{H}_J\, |2\rangle_j}{U} \notag\\
		&= \frac{4J^2}{U}.
	\end{align}
	
	Substituting these into $\hat{h}^{(2)}$ yields
	\begin{equation}
		\hat{h}^{(2)} = \sum_{j} \frac{2J^2}{U} e^{-i\theta}\, |2\rangle_{j+1}\langle 2|_j + \text{H.c.} + \sum_j \frac{4J^2}{U}\, |2\rangle_j\langle 2|_j,
	\end{equation}
	where the first term describes doublon hopping that acquires a statistical phase $e^{-i\theta}$ per hop, and the last term is a uniform on-site energy shift $4J^2/U$, which only contributes an overall constant to the energy spectrum.
	
	Combining $\hat{h}^{(0)}$, $\hat{h}^{(1)} = 0$, and $\hat{h}^{(2)}$, the full effective Hamiltonian is
	\begin{equation}
		\hat{H}_{\text{eff}} = \left(U + \frac{4J^2}{U}\right)\sum_j |2\rangle_j\langle 2|_j + \sum_{j} \frac{2J^2}{U} e^{-i\theta}\, |2\rangle_{j+1}\langle 2|_j + \text{H.c.}
	\end{equation}
	Introducing the doublon operators $\hat{c}_j^\dagger |0\rangle = |2\rangle_j$ and dropping the overall on-site energy shift, which does not affect the dynamics, we obtain the effective doublon Hamiltonian
	\begin{equation}
		\hat{H}_{\text{d}} = \frac{2J^2}{U} \sum_{j} e^{-i\theta}\, \hat{c}_{j+1}^\dagger \hat{c}_j + \text{H.c.},
		\label{eq:H-eff-final}
	\end{equation}
	where $\hat{c}_j^\dagger$ ($\hat{c}_j$) creates (annihilates) a doublon at site $j$. Equation~\eqref{eq:H-eff-final} shows that the doublon acquires an effective hopping phase $-\theta$ per nearest-neighbor bond.
	
	For a three-site ring ($N=3$), setting $\theta = \pi/6$ precisely reproduces the ideal chiral-circulation Hamiltonian derived in Appendix~\ref{sec:S2}. The anyonic statistical angle thus provides a natural many-body mechanism for realizing perfect chiral circulation without external fluxes.
	
\end{document}